\pdfoutput=1
\PassOptionsToPackage{dvipsnames}{xcolor}

\documentclass[11pt]{article}

\usepackage[final]{acl}

\usepackage{times}
\usepackage{latexsym}

\usepackage[T1]{fontenc}

\usepackage[utf8]{inputenc}

\usepackage{microtype}

\usepackage{inconsolata}

\usepackage{graphicx}

\usepackage{svg}
\usepackage{xurl}
\usepackage{afterpage}
\usepackage{booktabs}
\usepackage{amsmath}

\title{LLM generated responses to mitigate the impact of hate speech}

\author{
 \textbf{Jakub Podolak\textsuperscript{1}},
 \textbf{Szymon Łukasik\textsuperscript{2}},
 \textbf{Paweł Balawender\textsuperscript{2}},
 \textbf{Jan Ossowski\textsuperscript{2}},
\\
 \textbf{Jan Piotrowski\textsuperscript{2,5}},
 \textbf{Katarzyna Bąkowicz\textsuperscript{3}},
 \textbf{Piotr Sankowski\textsuperscript{2,4,5}}
\\
\\
 \textsuperscript{1}University of Amsterdam,
 \textsuperscript{2}University of Warsaw,
 \textsuperscript{3}SWPS University,
 \\
 \textsuperscript{4}IDEAS NCBR,
 \textsuperscript{5}MIM Solutions
\\
 \small{
   \textbf{Correspondence:} \href{mailto:jakub.podolak@student.uva.nl}{jakub.podolak@student.uva.nl},
    \href{mailto:sp.lukasik@student.uw.edu.pl}{sp.lukasik@student.uw.edu.pl},
   \href{mailto:sank@mimuw.edu.pl}{sank@mimuw.edu.pl}
 }
}

\begin{document}
\maketitle



\begin{abstract}
In this study, we explore the use of Large Language Models (LLMs) to counteract hate speech. We conducted the first real-life A/B test assessing the effectiveness of LLM-generated counter-speech. During the experiment, we posted 753 automatically generated responses aimed at reducing user engagement under tweets that contained hate speech toward Ukrainian refugees in Poland.

Our work shows that interventions with LLM-generated responses significantly decrease user engagement, particularly for original tweets with at least ten views, reducing it by over 20\%. This paper outlines the design of our automatic moderation system, proposes a simple metric for measuring user engagement and details the methodology of conducting such an experiment. We discuss the ethical considerations and challenges in deploying generative AI for discourse moderation.

\end{abstract}

\afterpage{
    \begin{figure*}[t]
      \includegraphics[width=2\columnwidth]{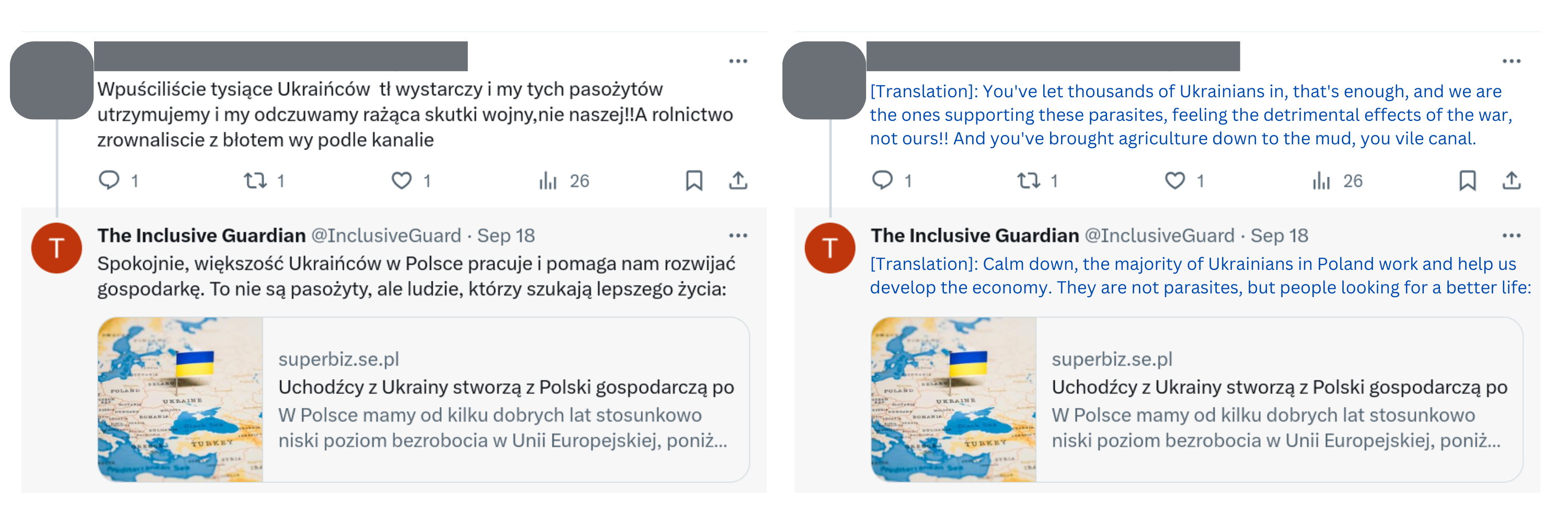}
      \caption{Example of detected, harmful tweet and our bot's counter-speech intervention conducted during the experiment. The original is on the left, and the translation to English is on the right.}
      \label{fig:example_intervention}
    \end{figure*}
}

\section{Introduction}
The full-scale conflict in Ukraine that started in February 2022 resulted in a humanitarian crisis and forced millions of people to leave their homes. In neighboring Poland, the conflict, its financial aspects, and help towards refugees were heavily discussed on social media platforms such as X (previously Twitter). In such a conversational environment, hate speech is used as a tool to reinforce the narratives on each side. It also contributes to the propagation of disinformation by distancing it from facts and highlighting the emotional aspect. Due to its strong impact on attitudes and beliefs, hate speech requires neutralization, which should take place as soon as possible after publication.

We define hate speech on the guidelines outlined in the Resolution of the Parliamentary Assembly of the Council of Europe (COE) \cite{coe2021resolution}. According to these, hate speech includes any form of expression that spreads, incites, promotes, or justifies hatred based on various forms of intolerance. This helps us identify behaviors that aim to dehumanize individuals, making them susceptible to stigmatization, discrimination, and violence.


In this work, we outline the design of our automatic detection and response system, describe the experiment's methodology, analyze and provide commentary on the results, and share additional insights that may be valuable for future work in automatic moderation systems. Furthermore, we discuss the ethical aspects of such interventions. Notably, we define a new Engagement metric that can be useful for counter-speech experiments. To our knowledge, this is the first experiment using A/B tests to demonstrate the effectiveness of counter-speech interventions on Twitter.

Our primary contributions in this work include:
\begin{enumerate}
\itemsep0em 
\item Developing a harmful tweet detection model specifically tailored for identifying hate speech against Ukrainian refugees.
\item Creating an intervention system using a Retrieval-Augmented Generation (RAG) approach.
\item Introducing a novel Engagement metric to measure the effectiveness of counter-speech interventions in reducing user engagement with harmful content.
\item Conducting the first live A/B test on Twitter, demonstrating the effectiveness of counter-speech interventions using generative AI.
\item Analyzing the results, and providing insights for future automatic moderation systems.
\end{enumerate}

\section{Related Work}
\paragraph{Hate Speech Detection}
Detecting hate speech on Twitter is a highly significant problem that has been studied by numerous research groups, see for example~\cite{8292838,10.1007/978-981-19-7455-7_17}. Deep learning techniques for this task have also been explored in~\cite{Badjatiya}. With the breakthrough of introducing the Transformer architecture~\cite{transformer} and Large Language Models like BERT~\cite{devlin2019bert}, the work \cite{Mozafari} has demonstrated that using them for this problem yields satisfying results. Moreover, BERT models enable the utilization of transfer-learning \cite{Zhuang} and fine-tuning techniques to adapt models for new tasks with limited data \cite{ALI2022101365} \cite{Tida}. With the availability of large RoBERTa models trained on Polish language data \cite{dadas2020pretraining}, we can leverage transfer learning to develop our own hate speech classifier in Polish.



\paragraph{LLM Generated Responses}
Generative models, especially GPT models \cite{gpt2}, have proved to perform well in different generative tasks without any additional training or fine-tuning, just by giving a couple of examples in prompt \cite{gpt3}. This technique of adapting LLMs for new tasks by demonstrating examples is referred to as few-shot learning or in-context learning \cite{incontext}. Wang et al. show \cite{wang2023evaluating} that by using in-context learning, the GPT-3 \cite{gpt3} model can generate high-quality, informative, and persuasive explanations of why the given text is hateful. Authors identify the potential of GPT-3 as a valuable tool for combating hate speech. On the other hand, the GPT models can generate biased \cite{gpt3bias} or hallucinated \cite{hallucination} content, and the generated texts should be approached with caution.

The usage of GPT models for content moderation online has already been researched, for example, by Axelsen et al. \cite{axelsen}. The authors of this work utilize GPT-3.5 models to identify and report toxic content and to reward positive contributions in messages among people. The method proposed in our work goes further by generating the responses that directly fight toxic content.  Russo et al. \cite{russo2023countering} show that with a source of knowledge and correct prompt, LLMs indeed can generate countering, emotional responses to fake, misleading, or manipulative content online. In a related study conducted concurrently with our experiment, a similar study generating counter-speech using Retrieval-Augmented Generation was conducted \cite{WarOfWords}. However, contrary to our paper, neither of these performs live experiments that verify the efficacy of using LLMs to change the discourse in real social media. 

\paragraph{Counter-speech}
Counter-speech has been highlighted by platforms like Facebook as an effective long-term strategy to combat hate speech. Facebook publicly stated that counter-speech is a critical tool that can be more impactful over time compared to simply removing offensive content \cite{Bartlett2015}.

Several studies have examined the nature of counter-speech, like linguistics aspects \cite{ThouShaltNotHate}, hate and counter users characteristics \citet{mathew2018analyzing}, and strategies \cite{Counterusersinteraction}.

Despite these findings, there is limited research on how counter-speech influences network dynamics over time. We found only one study that addressed this gap by showing that counter-speech can reduce the likelihood of neighboring users adopting hate speech, thus dampening the spread of negativity within social networks \cite{CounterSpeechPostHocAnalysisPaper}.




\section{Dataset}
\paragraph{Tweets Collection}
To obtain tweets, we used the Twitter v2 API with a broad query focused on the Russo-Ukrainian War. From March 3, 2022, to January 24, 2023, we collected approximately 2 million tweets. However, only a small fraction discussed our specific topic — assistance to Ukraine. To address this, we refined our query to focus on humanitarian aid, financial assistance, and refugee rights, resulting in an additional 215,000 relevant tweets. In total, we compiled approximately 2,215,000 tweets, referred to as the Twitter dataset. All tweets are in Polish and filtered using Twitter's language parameter. Details of both queries can be seen in \autoref{appendixA}.

\paragraph{Initial Dataset}
The initial dataset was created manually by reviewing random samples from the Twitter dataset. Initially, it consisted of 20 tweets classified as harmful, according to the hate speech definition we followed in this work. We were able to find 20 more examples by examining other tweets of the authors of the first 20 positives, effectively doubling our training set. This dataset provided a foundation for initial model training and subsequent active learning steps, although it was still insufficient to fully train the Hate RoBERTa classifier (described in \autoref{models_section}).

\paragraph{Dataset Expansion}
To expand the dataset, we adopted an iterative approach. In each iteration, we selected a batch of 1000 new tweets collected within the last day. Our model assigned a prediction score to each tweet, helping us manually select the most representative examples to add to the dataset. We focused on both positive (containing hate speech towards Ukrainian migrants in Poland) and negative (not containing hate speech) examples. In total, we conducted six iterations (between December 2022 and February 2023), obtaining the final dataset of 632 tweets, including 211 positive examples.

\paragraph{Annotation Process and bias}
Not all calls for suspending aid are motivated by hate; some may be based on concerns regarding the efficiency, direction, or implications of aid or on political, strategic, or ethical considerations that do not dehumanize or stigmatize a population. However, the tweets we categorized as hate speech share specific characteristics: they dehumanize the Ukrainian population and use language that incites fear, spreads misinformation, and promotes division between refugees and the host society. We recognize that distinguishing between criticism and hate speech can be subjective, leading to bias. As authors, we disclose that we live in a city with many refugees from Ukraine, which may have influenced our perspective.

\paragraph{Verified articles}
\label{verified_articles_subsection}
In order to make our responses more credible, we enhanced our dataset with articles from government institutions, private organizations, Polish press outlets, and other trusted sources. We verified the credibility of each article and maintained a database of 23 articles, along with their text and URLs. We continually updated the database and manually created article summaries to aid prompt construction. You can find the list of articles in section \autoref{sec:verified_articles}.


\section{Hate Classifier Model}
\label{models_section}

The Hate Classifier is designed to detect specific types of hate speech against aid for Ukraine, as defined in our dataset. Given the iterative process described in the Dataset section, this model leverages a curated set of tweets. We employed two primary approaches for classification:
\begin{enumerate}
\itemsep0em 
    \item \textbf{Hate RoBERTa - Fine-Tuning Approach:} 
    We fine-tuned an open-source RoBERTa model pre-trained on a Polish dataset\footnote{Polish RoBERTa Base model on Hugging Face: \url{https://huggingface.co/sdadas/polish-roberta-base-v2}}~\cite{dadas2020pretraining}. This method used transfer learning to adapt the model to our dataset. 
    \item \textbf{Embeddings + Logistic Regression Approach:}
    Additionally, we investigated a more streamlined approach, utilizing high-quality word embeddings from the text-embedding-ada-002 model via the OpenAI API to train a logistic regression classifier. It allows for efficient model updates. Notably, it proved to be a practical solution for frequent updates and continuous usage scenarios.
\end{enumerate}

The results of the models can be found in \autoref{tab:performance_metrics}. We did not compare them with other models for detecting hate speech due to the topic-specific nature of the problem. Our models are tailored to detect hate speech related to calls for aid for Ukraine. Our primary objective in this study is to assess the impact of our system on engagement through A/B testing. As a result, our aim was to develop a model with maximum precision.

\begin{table}[h!]
    \centering
    \small
    \begin{tabular}{lccc}
        \toprule
        \textbf{Method} & \textbf{Accuracy} & \textbf{Precision} & \textbf{Recall} \\ 
        \midrule
        Hate RoBERTa & 0.83 & 0.75 & 0.70 \\
        Embeddings + LR & 0.80 & 1.00 & 0.64 \\
        \bottomrule
    \end{tabular}
    \caption{Performances of Hate RoBERTa fine-tuning and Embeddings + Logistic Regression approaches.}
    \label{tab:performance_metrics}
\end{table}

\section{Retrieval-Augmented Generation}
\label{section:RAG}
To generate factual and reliable responses to harmful tweets, we implemented a Retrieval-Augmented Generation (RAG) approach \cite{ragDBLP:journals/corr/abs-2005-11401}. This method enhances the model's responses by incorporating relevant, verified information from external sources.

\paragraph{Vector Embeddings of Articles}
As described in \autoref{verified_articles_subsection}, we maintained a database of 23 verified articles from reputable sources which were embedded using the text-embedding-ada-002 model. When a harmful tweet was detected, we calculated its relevance to each article in our database using cosine similarity. The top three relevant articles were selected and included in the GPT prompt, ensuring that the responses referenced specific, verified information. Additionally, few-shot learning was utilized to achieve higher quality \cite{gpt3}.

\paragraph{Response Generation with GPT-4}
To generate responses, we initially evaluated several models, including GPT-3.5 and fine-tuned GPT-2 models. We conducted a survey to compare these models (details in \autoref{appendix:survey_results}). However, upon the release of GPT-4, we decided to switch to it, basing our decision on other researchers' findings that demonstrated its superiority over previous models on NLP and reasoning benchmarks. The prompt included instructions to generate responses in a Twitter-appropriate style, a brief description of the situation in Ukraine, summaries of the most relevant news articles and their sources, and example interactions (few-shot). For the counter-speech examples provided in the prompt, we chose two that seemed most relevant to the issue and that effectively countered the original hate speech. These examples were fixed and remained the same for each response generation. The shortened prompt can be found in \autoref{fig:shortened_prompt}, and the complete prompt can be found in \autoref{appendixUsedPrompts}. We enforced a 200-character limit on the generated content.

\begin{figure}[h]
  \centering
  \begin{minipage}{0.8\linewidth}
    \textcolor{blue}{
[Translation]: You are a Twitter user, a Polish person who adheres to humanitarian values and believes in helping others.
Your task is to engage in conversations on Twitter and combat hateful or false content propagated by other users. 
}
    \textcolor{red}{top 3 relevant verified articles texts and links}
    \textcolor{orange}{2 examples of tweets and positive responses}
  \end{minipage}
  \caption{Shortened prompt with RAG and few-shot for response generation. The full prompt can be found in \autoref{appendixUsedPrompts}}
  \label{fig:shortened_prompt}
\end{figure}

\paragraph{Cost Considerations}
The cost of generating responses was a significant factor in our methodology. While the cost of generating embeddings for tweets and articles was negligible, the cost of using the GPT-4 model for response generation was calculated to be \$0.048 per tweet. For more details, refer to \autoref{appendix:cost_calculations}.

\paragraph{Evaluation of Generated Responses}
To guarantee that we do not post harmful content ourselves, every reply generated by our system was verified manually before posting. In \autoref{subsection:tweet-posting}, we specify how we conducted the experiment, including how the manual verification was done. 

\section{Experiment}
The main hypothesis of this study is that automatically generated counter-speech can effectively decrease user engagement with harmful tweets, making them less convincing to users. We introduced a novel Engagement metric to validate this hypothesis and conducted an A/B test to determine its effectiveness. We aimed to assess whether the mean Engagement value is significantly lower in the Experimental Group (with model intervention) compared to the Control Group.

\begin{enumerate}
\itemsep0em 
    \item \textbf{Main Hypothesis:} Automatically generated counter-speech will decrease user engagement with harmful tweets.
    \item \textbf{Secondary Hypothesis:} Replies generated with verified links will increase the ratio of replies to impressions compared to replies without links.
\end{enumerate}

\paragraph{Definition of Engagement Metric}
\label{subsection:engagement-metric}
We define engagement as the ratio of the change in likes to the change in impressions over a given period. This metric is designed to capture user interaction relative to a tweet's exposure. Unlike traditional metrics, our engagement definition incorporates the tweet's visibility, providing a more nuanced understanding of user interaction.

\[
E(i) = \frac{\Delta \text{likes}(i)}{\Delta \text{impressions}(i)}
\]

where $\Delta \text{likes}(i)$ and $\Delta \text{impressions}(i)$ represent the change in likes and impressions, respectively, for tweet $i$ during the monitoring period. Viewing tweet impressions (a measure of a tweet's visibility) has been available since December 22, 2022 \cite{TwitterSupport2022}.

\paragraph{Experimental Setup and Data Collection}
\label{subsection:experiment-setup}
We utilized the hate speech detection and response generation systems described above. A Twitter bot account was created to fetch tweets related to Ukrainian refugees in Poland using the Twitter v2 API. These tweets were classified four times daily: 10 AM, 2 PM, 6 PM, and 10 PM. We only retrieved tweets posted within the past four hours to ensure our responses could still impact tweet popularity.

Analyzing Twitter activity in Poland, we observed a significant drop in engagement between midnight and 6 AM, prompting us to conduct interventions only during the day. We assumed that the experiment's date would not affect the results, as our primary interest was the growth dynamics of engagement metrics rather than their absolute values. Classified responses were randomly assigned to the Experimental or Control groups with equal probability.

\paragraph{Tweet posting}
\label{subsection:tweet-posting}
In the Experimental Group, responses were generated using the model from \autoref{section:RAG} and sent to a human verifier for approval within 45 minutes. Out of 1,282 generated responses, we manually rejected (i.e. restrain from posting) 406 replies. The most common reason was the original tweet being wrongly classified as harmful (a false-positive), which accounted for 238 rejections. Other reasons were the generated response being of very low quality, off-topic, controversial or containing hallucinations. The specifics of the rejection criteria and the exact statistics can be found in \autoref{appendix:ModelAnswersAnalysis}.

\paragraph{Final statistics}
During the response period (from 24.08.2023 to 19.09.2023), we retrieved 61,507 tweets, of which 3,143 were classified as harmful by our model. After manual verification conducted by four reviewers, 729 tweets remained in the Control Group, and we posted 753 replies to harmful tweets in the Experimental Group. An example of our intervention is shown in \autoref{fig:example_intervention}. Detailed information regarding the query to the Twitter API, prompts, and further reproducibility aspects can be found in \autoref{appendix:experiment-setup-details}.


\section{Results}
Detected harmful tweets can be the roots of discussions (the original tweet) or replies to another existing tweet (not the original tweet). In the next sections, we will validate our hypotheses in both groups as the results differ. 

\paragraph{Engagement change}

Using Engagement metric from \autoref{subsection:engagement-metric}, we measure the engagement of tweet $i$ as follows:
$$
E(i) = \frac{\text{likes'}(i) - \text{likes}(i)}{\text{impressions'}(i) - \text{impressions}(i)}
$$
here $\text{likes}(i)$ and $\text{impressions}(i)$ represent the number of likes and impressions, respectively, of harmful tweet $i$ just after detection and intervention (possible 15-minute delay). The $\text{likes'}(i)$ and $\text{impressions'}(i)$ values are the likes and impressions at the end of the monitoring period (6 days after the last detection and intervention).

\begin{figure}[!h]
  \includegraphics[width=0.85\columnwidth]{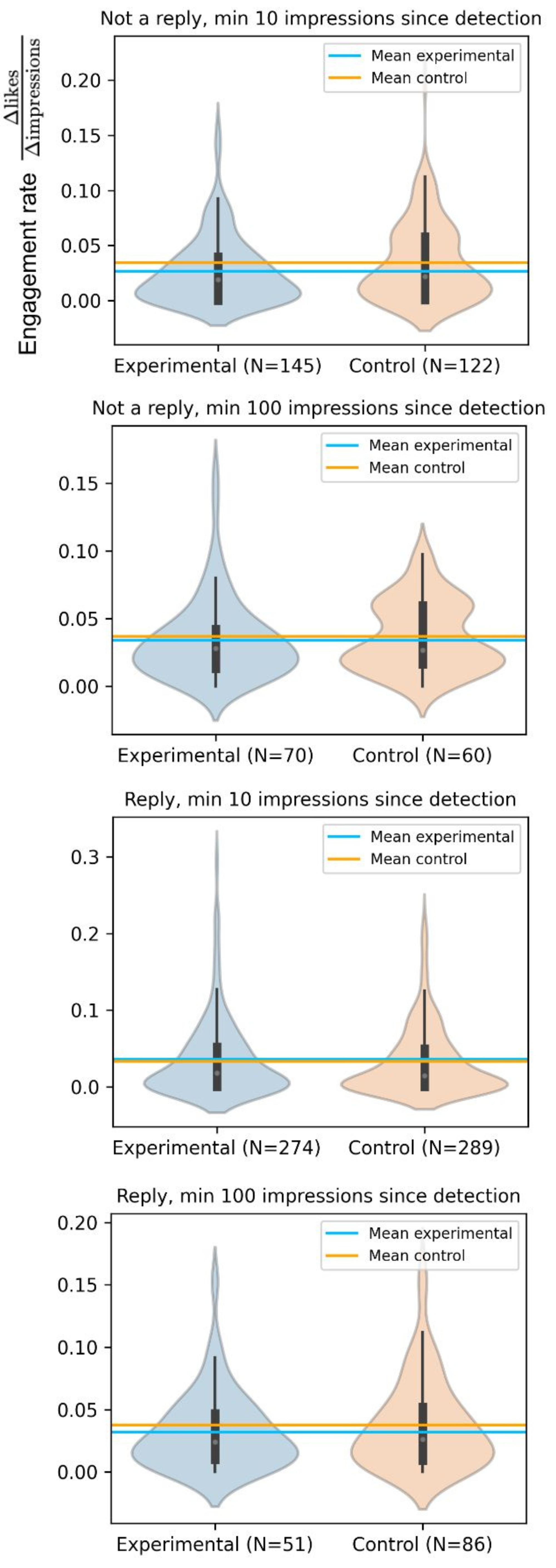}
  \caption{Distribution of engagement rate calculated as $\frac{\Delta \text{likes}}{\Delta \text{impressions}}$ across harmful tweets in the Experiment and Control group, depending on the data subset}
  \label{fig:distribution_engagement_rate}
\end{figure}

To validate our hypothesis, we aim to verify if the mean $E$ value is significantly lower in the Experimental Group compared to the Control Group. Furthermore, to ensure accurate calculation of the $E$ value, we must select the minimum value for the $\Delta \text{impressions}(i)$. We evaluated the impact of selecting minimum thresholds of 10 and 100 impressions. These four sets are visualized in \autoref{fig:distribution_engagement_rate}. We excluded tweets with anomalies, where the final number of likes was lower than the initial count (when someone unliked the post during the experiment period, representing 1\% of the collected data).

\begin{table}
    \centering
    \small

    \begin{tabular}
    {p{0.18\linewidth} | p{0.1\linewidth} | p{0.1\linewidth} | p{0.1\linewidth} | p{0.09\linewidth} | p{0.08\linewidth} }
        \hline
        Harmful tweet type & CG mean & EG mean & Diff (\%CG) & p (t-test) & p (Bstr) \\
        \hline
        Original, min 10 $\Delta$impr & 0.0346 & 0.0266 & -23\% & 0.027 & 0.003 \\
        \hline
        Original, min 100 $\Delta$impr & 0.0365 & 0.0337 & -8\% & 0.279 & 0.175 \\
        \hline
        Reply, min 10 $\Delta$impr & 0.0331 & 0.0362 & +8.3\% & 0.775 & 0.865 \\
        \hline
        Reply, min 100 $\Delta$impr & 0.0379 & 0.0319 & -16\% & 0.148 & 0.106 \\
        \hline
    \end{tabular}
    \caption{Mean Engagement and difference between Control (CG) and Experimental (EG) groups, depending on whether the tweet we answered to was original or a reply. (Diff - Difference between means, Bstr. - Bootstrapping)}
    \label{fig:engagement_table}
\end{table}

We observe that in three of these four scenarios, the mean engagement is lower in the Experimental group \autoref{fig:engagement_table} than in the Control group. That's not the case only when we reply to a harmful tweet that is already a reply to another tweet.

The most significant change can be seen when we reply to a harmful, original tweet with a minimum of 10 impressions in the monitoring period - In this group, we reduce the engagement by 23\%. To test the significance, we use the p-value obtained with the t-test. Also, we used the bootstrapping \cite{bootstrapping} technique to calculate the p-value. To do that, we sampled (with replacement) $10^4$ data samples from the Control group and compared the distribution of their means with our Experimental mean. Welch's t-test \cite{welch} gives a p-value < 0.05, and the bootstrapped p-value is < 0.01, making the result statistically significant.

It is worth noting that the lower engagement coefficient can also be a result of an increased number of impressions. We want to reduce the engagement of users without increasing the number of impressions of harmful tweets. Indeed, in the Experimental group, the median value ($0.5$ percentile) of both likes and impressions is lower compared to the Control group (\autoref{fig:likes_impressions_table}).

This significant, positive change is not observed when we respond to non-original tweets with fewer impressions (min 10). This is due to the cascading arrangement of comment content. A person joining the discussion may not notice entries in the cascades, but they will certainly notice the main comment segment, which will make neutralizing information reach them faster. Activating the bot in main discussions, as replies to original tweets, will, therefore, be more effective in neutralizing the social consequences of hate speech. We conclude that our intervention reduces the likes per impression coefficient for original tweets with min. 10 impressions during the experimental phase. This result is further discussed in \autoref{sec-discussion}.

\begin{table}
  \small
  \begin{tabular}{p{0.14\linewidth} | p{0.12\linewidth} | p{0.14\linewidth} | p{0.16\linewidth} | p{0.16\linewidth}}
    \hline
        Percentile & EG likes change & CG likes change & EG impr. change & CG impr. change \\
    \hline
    min & 0 & 0 & 10 & 10 \\
    \hline
    25\% & 0 & 0.25 & 27 & 27.25 \\
    \hline
    50\% & 1 & 3 & 78 & 87.5 \\
    \hline
    75\% & 14 & 59 & 606 & 1701 \\
    \hline
    max & 2067 & 2292 & 240651 & 173173 \\
    \hline
  \end{tabular}
  \caption{
    Percentile values of change in likes and impressions (impr.) in Experimental (EG) and Control (CG) groups. We report only original, harmful tweets with at least ten impressions gathered in the experiment.
  }
  \label{fig:likes_impressions_table}
\end{table}

\paragraph{Change in the number of replies}

As a result of further analysis of our data, we discovered that the ratio of the number of replies to a harmful tweet to its impressions is greater among the tweets that our bot interacted with. Similarly, as in the case of engagement, we measure the change in number of replies to the tweet $i$ as the:
$$
R(i) = \frac{\text{replies'}(i) - \text{replies}(i)}{\text{impressions'}(i) - \text{impressions}(i)}
$$
Where analogously, $\text{replies}(i)$ and $\text{impressions}(i)$ is the number of replies and impressions to the tweet $i$ after detection and $\text{replies'}(i)$ and $\text{impressions'}(i)$ are the number of replies and impressions at the end of the monitoring. This statistic does not count our bots' replies, so we measure only the change in other users' replies.

\begin{table}
    \centering
    \small
    \begin{tabular}{p{0.2\linewidth} | p{0.1\linewidth} | p{0.1\linewidth} | p{0.1\linewidth} | p{0.14\linewidth} | p{0.08\linewidth} }
        \hline
        Harmful tweets' set & CG mean & EG mean & Diff (\%CG) & p (t-test) & p (Bstr) \\
        \hline
        Root, min 10 $\Delta$impr. & 0.0032 & 0.0081 & +158\% & 0.0012 & 0.000 \\
        \hline
        Root, min 100 $\Delta$impr. & 0.0026 & 0.0034 & +28\% & 0.1234 & 0.011 \\
        \hline
        Reply, min 10 $\Delta$impr. & 0.0028 & 0.0134 & +387\% & 1.26e-9 & 0.000 \\
        \hline
        Reply, min 100 $\Delta$impr. & 0.0022 & 0.0030 & +40\% & 0.1399 & 0.081 \\
        \hline
    \end{tabular}
    \caption{Mean ratio of replies number change to impressions (impr.) change between Control (CG) and Experimental (EG) groups, depending on the data subset. (Diff - Difference between means, Bstr. - Bootstrapping)}
    \label{fig:reply_count_table}
\end{table}

We compare mean ratios in the Experimental and the Control groups by selecting our tweets with respect to minimum impression change and whether a harmful tweet was a reply or not. We also exclude tweets for which the number of replies decreased over a period of the monitoring (users deleted their responses) -- it is about $0.3\%$ of collected data.

In each of the four scenarios, the mean ratio increased. The increase is greater for tweets with a minimum impression change of 10 than for tweets with a minimum impression change of 100. The greatest increase, by $387 \%$, is observed when selecting non-original tweets with a minimum impression change of 10. 

We also tested the significance using bootstrapping and Welsch's t-test with the null hypothesis that the mean of the analyzed ratio is equal among groups. The results are presented in the \autoref{fig:reply_count_table}. The change seems to be significant only when selecting tweets with a minimum impression change of 10, with a bootstrapped p-value equal to zero \footnote{That means that every mean ratio of the tweets sampled from the Control Group was less than the mean ratio of the Experimental Group.}. When selecting tweets with a minimum impression change of 100, the change is not significant.
We conclude that our intervention increases the number of replies per impression for tweets with min. 10 impressions during the experimental phase. This result is further discussed in the Discussion section.

\paragraph{Verified links and their impact}
Let us now explore other interesting observations regarding the verified links in the tweets we found after the experiment. The results may serve as an insight for future automatic response generation systems.

Our model was prompted to support its responses with links from our base of verified sources whenever they applied to the subject of a tweet it responded to. Of the 753 responses posted during our experiment, 398 contained a link to one. Most of those responses consisted of a short statement pointing to a fact contradicting the original post and encouraged to check the source provided through a link at the end of the response.

One interesting observation is that when we answered original tweets (non-replies) with a response containing a link to a source, the increase of replies to the tweet was 24\% higher than in the case of responses that were not backed by a source. We observed this on the set of tweets with an impression change of at least 10. These results have a t-test p-value < 0.02 and bootstrap p-value < 0.001, which makes this finding even more interesting.

It could be hypothesized that this result has one of many causes: our method of preparing the knowledge base, our data sources considered controversial by X users, or distribution differences between tweets for which our bot posted a link versus tweets for which our bot didn't have a verified knowledge source. For example, tweets discussing financial aspects could have received more replies, and our bot posted more links when intervening with them. This could be a basis for further study.

Another observation we made was the lack of impact of links to sources on the changes to engagement (likes over impressions). One of our goals when providing links was to reduce user engagement in threads debunked by a verified source. Harmful tweets for which our bot replied with a link had a similar engagement ratio during the experiment as those for which our bot replied without a link. As before, this could have many causes and needs further research to give definitive results, but we do not conclude that verified knowledge helps reduce the popularity of harmful content.

\section{Discussion}
\label{sec-discussion}
This paper proposes the first automatic system based on the OpenAI GPT model for detecting hate speech on social media and counteracting it by posting automatically generated responses. The proposed system utilizes verified knowledge through fact-checked articles that the generative model can incorporate into intervention replies. We performed a live experiment with a proposed engagement metric that shows how counter-speech could be effective against hate speech.

We showed that harmful tweets that are not a reply and have gained at least 10 impressions during our experiment exhibit a 23\% lower Engagement rate (measured as likes over impressions) in the Experimental Group compared to the Control Group. This statistically significant effect is not observed for tweets that are replies, which typically receive fewer views than the original ones, or for tweets that have become very popular (gaining at least 100 impressions), where the dynamics of the engagement rate are harder to alter due to the presence of numerous other replies, which may dilute the impact of our reply. This suggests that automatic moderation systems such as ours perform better when engaged at the highest level of discussion and primarily with moderate popularity tweets, highlighting our interventions' nuanced effectiveness.

As an interesting result of our analysis, we present an effect of our replies on the ratio of change in replies to a harmful tweet to change in its impressions. We conclude that our replies increase this ratio, and we observe the greatest change for non-original tweets with a minimum impression count 10. These results may be explained by a social proof phenomenon, an idea that people are more likely to act if they see that others have already done so \cite{doi:10.1177/002200275800200106}.

We verified that at least $39\%$ of our replies to an original tweet with a minimum impression change of 10, and at least $48\%$ of our replies to a non-original tweet with a minimum impression change of 10 were the first reply to the tweet, thus initiating a discussion. It’s worth noting that despite these discussions, the number of impressions or popularity (measured as likes/impressions) does not increase. This may suggest that many additional responses come from the author or people already following the topic or discussion. However, we leave this analysis as part of further research.

We analyzed the impact of verified links on changes our responses made to the discussion and made a statistically significant observation of a 24\% higher increase in the response-to-impression ratio in responses supported by a source compared to those without one. This analysis was purely exploratory, so no concrete conclusions can be derived. However, we believe that further research on it could provide compelling results.

We also discussed the ethical aspects of such experiments and automated moderation systems. We are convinced that human supervision during the whole experiment, transparent information about our account being a bot, and anonymization of collected data ensure that the experiment was conducted in an informed and ethical manner. We believe this work can help develop better tools and systems for building safer online communities.

\section{Limitations}
The method and experiments presented in this paper have several limitations. Firstly, the experiment was conducted exclusively in the Polish language and focused solely on the war in Ukraine. While the LLM models used are multilingual and have demonstrated strong generalization capabilities, the behavior of social media users and experimental outcomes may vary significantly across different languages, communities, and discussion topics. Therefore, the generalizability of our results remains uncertain, and further research is necessary to evaluate the method's effectiveness in different linguistic and cultural contexts, social media platforms, and discussion topics.

Secondly, our method's reliance on human supervision is a significant limitation. Fully automated content moderation systems are prone to errors and can potentially spread misinformation. In this study, we did not assess our method's performance in a fully automated scenario, which limits our understanding of its applicability and reliability in real-world, unsupervised settings.

Additionally, our model only processes textual data, ignoring tweets with graphics, which are common in disinformation and hate speech. Integrating vision models, such as GPT-4o, which supports vision input, could enhance the capability to detect and respond to such multimedia content.

Furthermore, the engagement metric used in this study may not be suitable for popular accounts with large followings, as their ability to generate views could skew the metric, making it less effective for assessing the impact of counter-speech interventions.

Additionally, the popularity of a counter-speech profile could attract trolls who may counteract efforts by increasing the visibility of harmful posts. This underscores the importance of robust countermeasures and continuous monitoring to mitigate such risks.

Finally, the limitations of the detection and answering model, including potential misclassifications and hallucinations, pose significant ethical challenges. These errors can lead to confusion, frustration, and the spread of misinformation. Human oversight remains necessary to ensure automated responses' accuracy and appropriateness.

\section{Ethical Considerations}
\label{appendix:ethical}

\subsection{Acknowledgments}
We would like to acknowledge the use of ChatGPT by OpenAI for assisting in the drafting and editing of this paper. The tool was used to help with language polishing and to generate preliminary text, which was then thoroughly reviewed and edited by the authors. Any errors that remain are our own responsibility.

\subsection{Ethical Guidelines}
\label{ethical-considerations}
In developing our hate speech detection model, we used publicly available data from Twitter, ensuring data security and privacy. We followed Twitter's guidelines and informed the Ethics Committee of a major European university about our experiment and it was performed under their supervision. We manually verified bot-generated replies to prevent harm and provided clear information about the bot's identity. The collected data was anonymous, with only tweet IDs, public metrics, and text stored.

Twitter strictly prohibits any automated activities that encourage abuse, violence, hateful conduct, or harassment, both on and off our platform. However, the use of automation to counter hate speech is permitted within this context.\footnote{X policies page: \url{https://help.twitter.com/en/rules-and-policies/x-automation}}

\subsection{Ethical Impact and Broader Implications}
Addressing hate speech and disinformation on social media is a critical societal responsibility. The increasing volume and decreasing quality of information shared online necessitate automated tools for managing and moderating content. Our proposed system aims to detect and neutralize hate speech by generating counter-speech interventions. This approach can mitigate the negative effects of disinformation and promote a healthier information ecosystem.

However, the ethical implications of such a system must be carefully considered. The deployment of automated moderation tools requires stringent ethical oversight to ensure that they do not infringe on free speech or disproportionately target specific groups. Additionally, while the use of automation to combat hate speech is permitted within the context of Twitter's guidelines, it is crucial to maintain transparency and user awareness regarding the presence of bots.

The potential for misuse of such tools by unethical actors is another significant concern. Although intended for positive applications, these tools could be exploited for malicious purposes, such as generating false counter-speech to discredit true information. This highlights the need for responsible implementation and critical evaluation by researchers, policymakers, and industry professionals.

The increasing prevalence of machine-generated content on social media raises questions about the authenticity and quality of online discourse. Our approach contributes to this trend, potentially leading to a higher proportion of bot-generated content. The implications of this shift are not fully understood, but it is essential to consider the impact on user trust and the overall information ecosystem.

In conclusion, while our system presents a promising approach to addressing hate speech and disinformation on social media, its ethical implementation requires careful consideration of the potential risks and broader implications. Responsible deployment and ongoing evaluation are essential to maximize the benefits 
while minimizing the negative impacts.

\bibliography{base}

\clearpage
\appendix

\section{Reproducibility}
\label{appendixA}
In this section, we describe additional technical details that may be useful in reproducing the results obtained.

\paragraph{Harmful tweets fetching}
In order to fetch harmful tweets, we used elevated access to X v2 API and `GET /2/tweets/search/recent` endpoint through Tweepy python library. To construct the API query, we used a list of words, and hashtags that we considered as related. There is also the extra language mark "lang:pl" to ensure we fetch only tweets in Polish. The more broad query is described in \autoref{fig:wide_query_twitter} and more suitable to humanitarian aid for Ukraine is described in \autoref{fig:query_twitter}

\paragraph{Russo-Ukrainian War tweets fetching}
\begin{figure}[h]
  \centering
  \begin{minipage}{0.8\linewidth}
    \textcolor{blue}{'(ukraina OR ukraiński OR rosja OR rosyjski OR putin OR sowiecki OR kreml OR kremlowski OR mińsk OR NATO OR kijów OR moskwa OR zełeński OR sankcje OR rubel OR donbas OR UKR OR RUS OR \#ukraine OR \#ukraina OR \#russia OR \#rosja OR \#war OR \#wojna OR \#warinukraine OR \#wojnawukrainie OR \#wojnanaukrainie OR \#standwithukraine OR \#ukrainerussiawar OR \#putin OR \#ukrainewar OR \#putinwarcrimes OR \#ukraineunderattack OR \#russianaggression) -is:retweet lang:pl'}
  \end{minipage}
  \caption{Query used to search for tweets in the Twitter network}
  \label{fig:wide_query_twitter}
\end{figure}

\begin{figure}[h]
  \centering
  \begin{minipage}{0.8\linewidth}
    \textcolor{blue}{'(ukraina OR ukraiński OR ukraińcy OR ukraińców OR ukraińca OR bandera OR banderowcy OR banderowscy OR upadlina OR upadlińscy OR ukropol OR ukropolin OR wołyń OR wołyński OR wołyńskie OR ukrainizacja OR ukrainizacji OR ukrainizację OR przebywający OR pomoc OR dzicz OR ukry OR ukrowie OR przywileje OR dziczy OR wynocha OR pobór OR dezerter OR \#StopUkrainizacjiPolski OR \#ToNieNaszaWojna OR \#StopUkroPol OR \#StopbanderyzacjiPolski OR \#żebyPolskabyłapolska) -is:retweet lang:pl'}
  \end{minipage}
  \caption{Query used to search for harmful tweets in the Twitter network}
  \label{fig:query_twitter}
\end{figure}

\paragraph{Tweets classification}
In order to classify tweets, we used the `text-embedding-ada-002` embedding model offered by OpenAI API and the logistic regression model from the scikit-learn package. We calibrated the model using `sklearn.calibration.CalibratedClassifierCV` \cite{scikit-learn} to compensate for the unbalanced training classes. 

\paragraph{Response generation and verified links}
Links to verified articles and their text were stored in the online sheet tool, which allowed for the quick addition of new articles. The articles were fetched each time the responses were generated, and relevant ones were included in the system prompt of the generative model. As a generative model, we've chosen the `gpt-4` model accessed through OpenAI API. The prompt used to generate the responses is shown in the sections below.

\paragraph{Statistics and metrics of tweets}
In order to fetch statistics of harmful tweets and our replies, we used the `GET /2/tweets` endpoint of Twitter v2 API. 


\section{Experiment Setup Details}
\label{appendix:experiment-setup-details}

\subsection{Tweet Classification Timing}
Tweets were classified four times daily at 10 AM, 2 PM, 6 PM, and 10 PM from
24.08.2023 to 19.09.2023.

\subsection{Tweet Retrieval Delay}
We retrieved only tweets posted within the past four hours to ensure they were not saturated (i.e., reached more than 80\% of their final number of impressions). Our analysis indicated that after 12 hours, 80\% of tweets received 80\% of their final impressions, while less than 50\% reached this threshold within five hours. Therefore, we limited our response delay to four hours.

\subsection{Nighttime Activity Analysis}
By analyzing the difference in likes between consecutive 30-minute timestamps, we found that Twitter activity in Poland drastically drops between midnight and 6 AM. This motivated us to perform interventions exclusively during the day.

\subsection{Labeling Guidelines and Labeler Demographics}
We employed four labelers, all of whom were Polish, male, white, and aged between 20-25, with higher education backgrounds. To optimize the annotation process for topic classification dataset expansion, each tweet was assigned to a single labeler. The labelers were instructed to mark tweets that expressed hate speech towards Ukrainian migrants in Poland as positive examples and classify tweets that did not meet this criterion as negative examples. During manual verification of responses to tweets before intervention, labelers were instructed to exclude low-quality or off-topic responses, as well as those containing false or unverified information generated by a language model.

\subsection{Response Rejection Criteria}
Responses in the Experimental Group were rejected based on the following criteria:
\begin{itemize}
    \item The original tweet was not harmful (classifier false positive).
    \item The response was off-topic.
    \item The response was on-topic but low-quality.
    \item The quoted article did not match the content of the response.
    \item The response was potentially controversial.
    \item The model hallucinated by generating false statements.
\end{itemize}
Only the first criterion was checked for the Control Group.

\subsection{Detailed Process Flow}
\begin{figure}
    \centering
    \includegraphics[width=1\columnwidth]{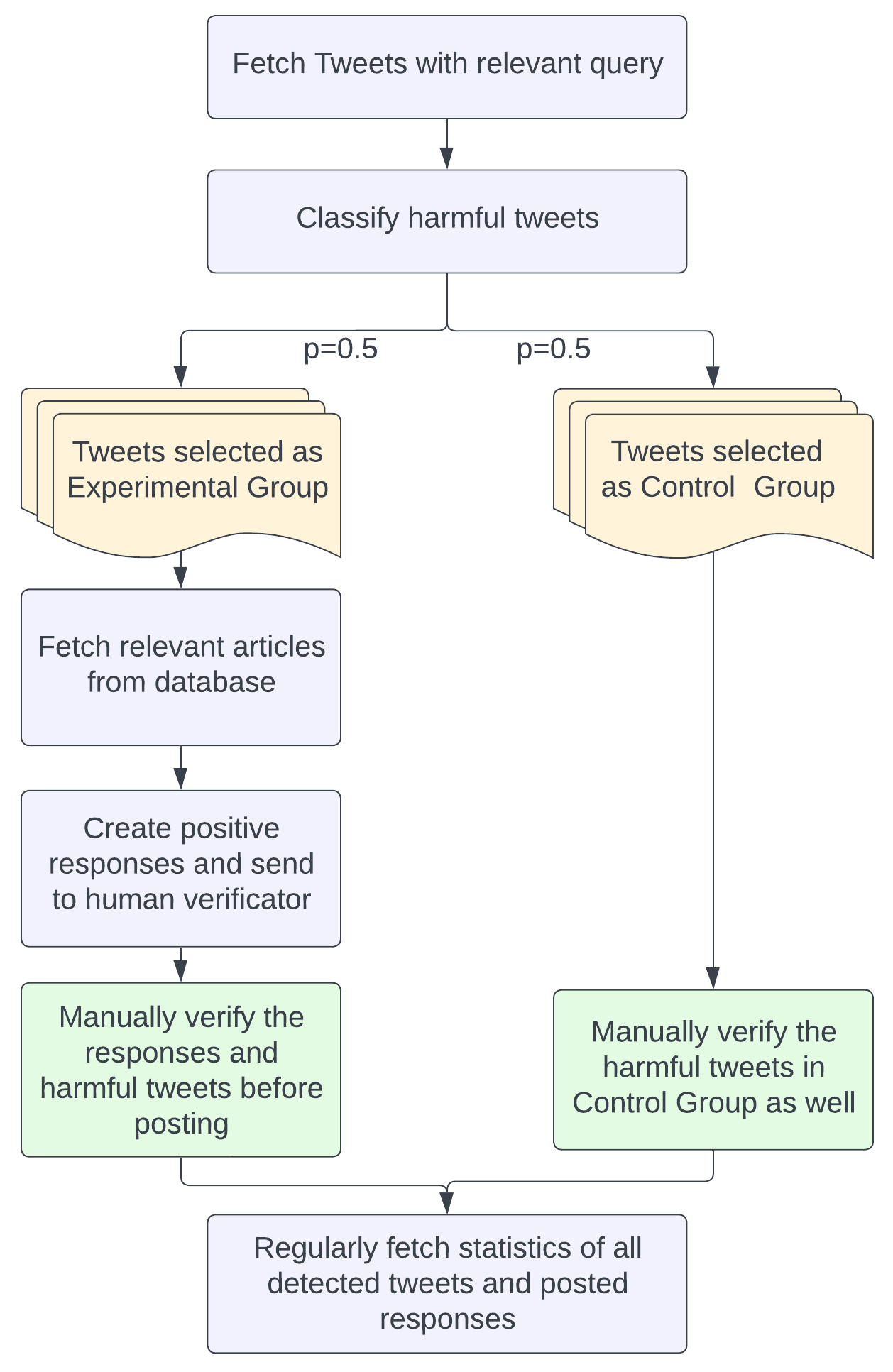}
    \caption{The high-level overview of proposed experiment algorithm}
    \label{fig:experiment_algorithm}
\end{figure}

The classified responses were randomly assigned to the Experimental or Control Group with equal probability. For the Experimental Group, responses were generated using verified knowledge and sent to a human verifier, who had 45 minutes to approve or reject them. If the generated tweet was not blocked, it was posted, and after 15 minutes, the first metrics (likes, impressions, replies) about both the harmful tweet and our bot's reply were fetched. The full process can be seen in \autoref{fig:experiment_algorithm}.

\subsection{Twitter Profile}
The bot account was named "The Inclusive Guardian" with the description: "I’m a bot dedicated to fostering neutral and empathetic discussions on Ukrainian refugees in Poland. Let’s combat hate together and promote a safe space!".\footnote{Bot account: \url{https://twitter.com/InclusiveGuard}}.

During the experiment, our Twitter profile had no followers and followed only the @NASA profile. We believe that this minimized any influence on the popularity of our replies. The primary audience for our replies were people who followed the accounts that posted the original harmful tweets, which we responded to.

\section{Verified Articles}
\label{sec:verified_articles}

We maintained a database of 23 articles, including their text and URLs, sourced from reputable news outlets. The articles were chosen based on their credibility, verified through independent sources, and their focus on specific issues related to the war in Ukraine and its impact on Poland. Below is the detailed list of these articles (titles were translated from Polish to English):

\begin{itemize}
\item \textbf{Title:} One-time Financial Benefit \
\textbf{Last Update:} 2023-07-28 \
\textbf{Category: Benefits and Allowances} \
\textbf{Source:} \url{https://ukraina.interwencjaprawna.pl/wp-content/uploads/2023/01/Broczura-SIP_PL.pdf}

\item \textbf{Title:} Child Benefit (so-called 500 Plus) \
\textbf{Last Update:} 2023-07-28 \
\textbf{Category: Benefits and Allowances} \
\textbf{Source:} \url{https://ukraina.interwencjaprawna.pl/wp-content/uploads/2023/01/Broczura-SIP_PL.pdf}

\item \textbf{Title:} Polish Authorities’ Expenditures on Aid to Ukraine \
\textbf{Last Update:} 2023-07-26 \
\textbf{Category: Expenditures and Costs} \
\textbf{Source:} \url{https://www.infor.pl/prawo/nowosci-prawne/5635962,Polska-pomoc-dla-Ukrainy-2022-ile-kosztowala.html}

\item \textbf{Title:} Number and Demographics of Ukrainians in Poland \
\textbf{Last Update:} 2023-07-26 \
\textbf{Category: Statistical Data} \
\textbf{Source:} \url{https://nbp.pl/wp-content/uploads/2023/04/Sytuacja-zyciowa-i-ekonomiczna-migrantow-z-Ukrainy-w-Polsce_raport-z-badania-2022.pdf}

\item \textbf{Title:} Ukrainian Army: Russians Want to Practice Destroying Civilian Ships Sailing to and from Ukrainian Ports [LIVE REPORT] \
\textbf{Last Update:} 2023-07-26 \
\textbf{Category: Military Activities} \
\textbf{Source:} \url{https://www.infor.pl/prawo/nowosci-prawne/5635962,Polska-pomoc-dla-Ukrainy-2022-ile-kosztowala.html}

\item \textbf{Title:} Sentiment and Opinion of Poles on Helping Ukrainians \
\textbf{Last Update:} 2023-07-26 \
\textbf{Category: Opinions and Analysis} \
\textbf{Source:} \url{https://www.rp.pl/spoleczenstwo/art38594711-polacy-wciaz-chca-pomagac-ukraincom-ale-na-duzo-mniejsza-skale-niz-zaraz-po-wybuchu-wojny}

\item \textbf{Title:} GDP Growth in Poland in the 1st Quarter of 2023 \
\textbf{Last Update:} 2023-07-26 \
\textbf{Category: Statistical Data} \
\textbf{Source:} \url{https://businessinsider.com.pl/gospodarka/wzrost-polskiej-gospodarki-najwyzszy-w-europie-w-pkb-gonimy-szwajcarie/zs7lr6l}

\item \textbf{Title:} Scandalous Banners at the Ekstraklasa Match. Hooligans Did Not Show Off - Śląsk Wrocław \
\textbf{Last Update:} 2023-07-30 \
\textbf{Category: Current News} \
\textbf{Source:} \url{https://sport.wprost.pl/pilka-nozna/ekstraklasa/11328016/skandaliczne-transparenty-na-meczu-ekstraklasy-kibole-sie-nie-popisali.html}

\item \textbf{Title:} Import of Grain, Food from Ukraine to the EU and Poland \
\textbf{Last Update:} 2023-04-11 \
\textbf{Category: Opinions and Analysis} \
\textbf{Source:} \url{https://www.osw.waw.pl/pl/publikacje/komentarze-osw/2023-04-11/wzrost-importu-zywnosci-z-ukrainy-do-ue-uwarunkowania-i}

\item \textbf{Title:} President Zelensky’s Thanks for Help \
\textbf{Last Update:} 2023-04-05 \
\textbf{Category: Current News} \
\textbf{Source:} \url{https://samorzad.pap.pl/kategoria/aktualnosci/prezydent-zelenski-drogi-rzeszowie-dziekuje-ci-za-ze-zostales-pierwszym}

\item \textbf{Title:} How Much Tax Do Ukrainians Pay in Poland? \
\textbf{Last Update:} 2022-11-01 \
\textbf{Category: Expenditures and Costs} \
\textbf{Source:} \url{https://www.tvp.info/64268612/ukraincy-zaplacili-w-polsce-10-mld-zlotych-podatku}

\item \textbf{Title:} Retirement After a Few Days of Work \
\textbf{Last Update:} 2023-06-16 \
\textbf{Category: Benefits and Allowances} \
\textbf{Source:} \url{https://demagog.org.pl/wypowiedzi/emerytura-dla-ukraincow-po-dniu-pracy-to-nieprawda/}

\item \textbf{Title:} Transport of Goods Between Ukraine and the EU. On What Terms Does It Operate? \
\textbf{Last Update:} 2023-04-18 \
\textbf{Category: Others} \
\textbf{Source:} \url{https://demagog.org.pl/wypowiedzi/przewoz-towarow-miedzy-ukraina-i-ue-na-jakich-zasadach-dziala/}

\item \textbf{Title:} Tax Obligations of Ukrainian Citizens Working in Poland \
\textbf{Last Update:} 2022-11-17 \
\textbf{Category: Expenditures and Costs} \
\textbf{Source:} \url{https://raczkowski.eu/prohr/blog/2022/ukrainiec-przebywajacy-w-polsce-ponad-183-dni-placi-pit-oraz-inne-obowiazki-podatkowe-obywateli-ukrainy-pracujacych-w-polsce-pro-hr-listopad-2022.html#:~:text=Obywatel%20Ukrainy%2C%20kt%C3%B3rego%20pobyt%20w,rezydentem%20podatkowym%2C%20czy%20te%C5%BC%20nie.}

\item \textbf{Title:} More and More Foreigners Covered by Social Insurance \
\textbf{Last Update:} 2022-08-09 \
\textbf{Category: Expenditures and Costs} \
\textbf{Source:} \url{https://www.zus.pl/-/coraz-wi%C4%99cej-cudzoziemc%C3%B3w-obj%C4%99tych-ubezpieczeniem-spo%C5%82ecznym}

\item \textbf{Title:} Ukrainian Expenditures Boost Polish Retail \
\textbf{Last Update:} 2022-10-03 \
\textbf{Category: Statistical Data} \
\textbf{Source:} \url{https://www.rp.pl/handel/art37169331-wydatki-ukraincow-zasilaja-polski-handel}

\item \textbf{Title:} Ukrainian Refugees Will Make Poland an Economic Powerhouse? \
\textbf{Last Update:} 2022-12-24 \
\textbf{Category: Opinions and Analysis} \
\textbf{Source:} \url{https://superbiz.se.pl/wiadomosci/uchodzcy-z-ukrainy-stworza-z-polski-gospodarcza-potege-rozmowa-aa-BxYg-6UV8-ciwZ.html}

\item \textbf{Title:} Ukrainians Are Starting Companies in Poland \
\textbf{Last Update:} 2023-08-09 \
\textbf{Category: Opinions and Analysis} \
\textbf{Source:} \url{https://edialog.media/2023/08/09/co-dziesiata-firma-w-polsce-ukrainska/}

\item \textbf{Title:} Budget Revenues from Taxes and ZUS \
\textbf{Last Update:} 2023-08-09 \
\textbf{Category: Statistical Data} \
\textbf{Source:} \url{https://edialog.media/2023/08/09/co-dziesiata-firma-w-polsce-ukrainska/}

\item \textbf{Title:} Ukrainian Refugees Have Found Jobs in the Polish Labor Market \
\textbf{Last Update:} 2023-02-12 \
\textbf{Category: Statistical Data} \
\textbf{Source:} \url{https://www.rp.pl/rynek-pracy/art37993771-ukrainscy-uchodzcy-odnalezli-sie-na-polskim-rynku-pracy}

\item \textbf{Title:} Ukrainian Children with Priority to Nursery? That’s Not True \
\textbf{Last Update:} 2023-08-28 \
\textbf{Category: Opinions and Analysis} \
\textbf{Source:} \url{https://demagog.org.pl/wypowiedzi/ukrainskie-dzieci-z-pierszenstwem-do-zlobka-to-nieprawda/}

\item \textbf{Title:} How Much Money Did Poland Get from the EU to Support Refugees? \
\textbf{Last Update:} 2023-02-28 \
\textbf{Category: Expenditures and Costs} \
\textbf{Source:} \url{https://demagog.org.pl/wypowiedzi/ile-pieniedzy-polska-dostala-od-ue-na-wsparcie-uchodzcow/}

\item \textbf{Title:} Is Poland Deporting Ukrainians? Repeat of Last Year’s Fake News \
\textbf{Last Update:} 2023-09-05 \
\textbf{Category: Opinions and Analysis} \
\textbf{Source:} \url{https://demagog.org.pl/fake_news/polska-deportuje-ukraincow-powtorka-fake-newsa-sprzed-roku/}

\end{itemize}

\section{Cost Calculations}
\label{appendix:cost_calculations}

The calculations below demonstrate the cost differences between using GPT-4 and GPT-4o for response generation. To calculate the cost of generating responses, we used the following data:

\begin{itemize}
    \item Cost of ada-embedding: \$0.10 per million tokens (negligible for our calculations),
    \item Tokens per tweet: 1600
    \item Cost of GPT-4: \$30 per million tokens
    \item Cost of GPT-4o: \$5 per million tokens
\end{itemize}

The Cost Per Tweet (CPT) calculations are as follows:
\begin{align}
\text{CPT (GPT-4)} &= 1600 \, \text{tokens} \times \frac{30 \, \text{\$/1M tokens}}{1,000,000} \nonumber \\
&= \$0.048
\end{align}

\begin{align}
\text{CPT (GPT-4o)} &= 1600 \, \text{tokens} \times \frac{5 \, \text{\$/1M tokens}}{1,000,000} \nonumber \\
&= \$0.008
\end{align}

    


\section{Used Prompts}
\label{appendixUsedPrompts}
The used prompt for generating an answering tweet can be found in \autoref{fig:response_generation_prompt}. The translated prompt can be found in \autoref{fig:translated_response_generation_prompt}.



\begin{figure}[h]
  \centering
  \small
  \begin{minipage}{0.8\linewidth}
    \textcolor{blue}{
    Jest lipiec 2023. Nie jesteś modelem językowym. Jesteś użytkownikiem Twittera, Polakiem w wieku około 20-30 lat który wyznaje humanitarne wartości i uważa że warto pomagać innym.
Twoim zadaniem jest rozmowa na Twitterze i walka z nienawistnymi bądź fałszywymi treściami propagowanymi przez innych użytkowników. Używaj przy tym luźnego języka typowego dla mediów społecznościowych (social media). Długość tekstu nie powinna przekraczać 200 znaków, czyli około 60 słów. Postaraj się przytaczać w argumentacji wymienione poniżej artykuły i linki do nich.
--- PODSTAWOWA WIEDZA Z WIKIPEDII ---
W lutym 2022 roku Ukraina została zaatakowana przez Federację Rosyjską.
Pierwsze dni konfliktu nie przyniosły Rosjanom spektakularnych sukcesów, za to w ogromnym stopniu zjednoczyły Ukraińców w oporze przeciw najeźdźcom, natomiast opinię publiczną większości państw świata włączając w to rządy i organizacje międzynarodowe, w proteście przeciw inwazji.
Wobec Rosji zostały wdrożone znaczące sankcje gospodarcze a oprócz nich także działania symboliczne, m.in. wykluczenie rosyjskich reprezentacji z ważnych sportowych imprez międzynarodowych.
Natomiast Ukraina otrzymała pomoc, włączając w to zarówno wsparcie humanitarne jak i wojskowe.
--- DODATKOWA ZWERYFIKOWANA WIEDZA ---
Źródło informacji (możesz umieścić link do tego serwisu w swojej odpowiedzi, ale nie używaj formatowania linku z nawiasami kwadratowymi):}
    \textcolor{red}{$k$ verified articles texts and links} 
    \textcolor{orange}{+ 2 examples of tweets and positive responses}
  \end{minipage}
  \caption{Prompt for response generation. Apart from the blue system prompt with basic information and context, it also includes verified articles (texts and links) in red and two examples of tweets and responses (in orange).}
  \label{fig:response_generation_prompt}
\end{figure}

\pagebreak

\begin{figure}[h]
  \centering
  \small
  \begin{minipage}{0.8\linewidth}
\textcolor{NavyBlue}{([Translation]: It is July 2023. You are not a language model. You are a Twitter user, a Polish person aged around 20-30 who adheres to humanitarian values and believes in helping others.
Your task is to engage in conversations on Twitter and combat hateful or false content propagated by other users. Use a casual tone typical of social media. The text length should not exceed 200 characters, which is approximately 60 words. Try to reference the articles and links listed below in your arguments.
--- BASIC KNOWLEDGE FROM WIKIPEDIA ---
In February 2022, Ukraine was attacked by the Russian Federation.
The initial days of the conflict did not bring spectacular successes for the Russians, but they significantly united Ukrainians in their resistance against the invaders and garnered the support of the public opinion in most countries worldwide, including governments and international organizations, in protest against the invasion.
Significant economic sanctions were imposed on Russia, along with symbolic actions, such as the exclusion of Russian teams from major international sports events.
Meanwhile, Ukraine received aid, including both humanitarian and military support.
--- ADDITIONAL VERIFIED KNOWLEDGE ---
Source of information (you can include a link to this source in your response, but do not use link formatting with square brackets):)
}
    \textcolor{red}{$k$ verified articles texts and links}
    
    \textcolor{orange}{+ 2 examples of tweets and positive responses}
  \end{minipage}
  \caption{Translated prompt for response generation. Apart from the blue system prompt with basic information and context, it also includes verified articles (texts and links) in red and two examples of tweets and responses (in orange).}
  \label{fig:translated_response_generation_prompt}
\end{figure}

\section{Survey Results}
\label{appendix:survey_results}


\begin{table}[h!]
    \centering
    \small
    \begin{tabular}{p{0.25\linewidth}| p{0.12\linewidth} | p{0.12\linewidth} | p{0.12\linewidth}| p{0.12\linewidth} }
        \toprule
        \textbf{Feature} & \textbf{GPT-3 Davinci} & \textbf{GPT-3.5} & \textbf{GPT-2 Medium} & \textbf{GPT-2 XL} \\
        \midrule
        Pro-Ukrainian & 44.10\% & 84.24\% & 61.67\% & 75.56\% \\
        On-topic & 43.59\% & 86.67\% & 27.22\% & 52.22\% \\
        Comprehensible & 50.77\% & 85.45\% & 50.00\% & 69.44\% \\
        Written by a bot & 39.49\% & 23.03\% & 50.00\% & 37.78\% \\
        \bottomrule
    \end{tabular}
    \caption{Survey results from 12 respondents evaluating different response generation methods. The GPT-2 models used were \textbf{Polish GPT-2 Medium} \url{https://huggingface.co/sdadas/polish-gpt2-medium} and \textbf{Polish GPT-2 XL} \url{https://huggingface.co/sdadas/polish-gpt2-xl}.}
    \label{fig:form_table}
\end{table}

\section{Model Answers Analysis}
\label{appendix:ModelAnswersAnalysis}
In \autoref{subsection:experiment-setup}, we outline the process of our experiment, including the manual verification mechanism. Here we present a more detailed analysis of the rejected responses. Out of 1,282 generated responses, 406 were manually rejected. The reasons for rejections and their frequencies were as follows:

\begin{itemize}
    \item The original tweet was not harmful (false positive): 238 instances
    \item Low-quality response (response is in the topic but it is essentially a spam content): 91 instances
    \item Off-topic answer: 43 instances
    \item False or unverified information (hallucinated): 16 instances
    \item The article does not contain proper information: 15 instances
    \item Response was controversial (LLM answered inappropriately): 3 instances
\end{itemize}

\subsection{Detailed Analysis of Rejections}

\begin{itemize}
    \item \textbf{False Positives}:
    Many tweets were detected as harmful due to new events that were not included in our training set. An example was the grain crisis between Ukraine and Poland, where most comments were complaints rather than hate speech. This highlights the dynamic nature of geopolitical relations and the challenge of analyzing topics not covered in the training set.

    \item \textbf{Low-Quality Responses}:
    In 91 instances, the responses were relevant to the topic but not counter-speech. This suggests that the model needs further refinement to generate more persuasive content.

    \item \textbf{Off-Topic Answers}:
    Off-topic answers occurred 43 times. This often happened when the model encountered tweets with videos, images, or external links, which it couldn't analyze effectively. Additionally, the model struggled to respond appropriately to descriptions of videos or interviews that contained provocative statements designed to attract viewers, which were further debunked but did not constitute hate speech.

    \item \textbf{Hallucinations}:
    In 16 responses, the model hallucinated or provided false statements. It especially produced categorical statements and wanted to debunk true statements if they were even slightly criticized Ukrainian.

    \item \textbf{Mismatched Articles}:
    In 15 instances, the response quoted articles that did not contain relevant information. This issue arose particularly when responding to tweets referencing other users or involving complex narratives.

    \item \textbf{Controversial Responses}:
    In 3 instances, the model generated controversial responses. For example, it used the controversial statement "Ukrainians are our workforce" to affirm their contributions, which could be perceived as offensive.
\end{itemize}

\end{document}